\documentclass{llncs}

\usepackage[english]{babel}
\usepackage[utf8]{inputenc}
\usepackage[T1]{fontenc}

\usepackage{tikz}
\usetikzlibrary{arrows}
\usetikzlibrary{topaths}
\usetikzlibrary{calc}
\usetikzlibrary{shapes}

\usepackage{cleveref}
\renewcommand\cref[1]{\Cref{#1}}

\newcounter{mnotei}
\usepackage{listings}
\usepackage{xspace}

\newcommand{\clpfd}{CLP($\mathcal{FD}$)\xspace}
\newcommand{\fd}{$\mathcal{FD}$\xspace}

\title{The Ciao \clpfd Library \\ A Modular CLP Extension for Prolog}
\subtitle{(System Description)}

\author{%
  Emilio Jesús Gallego Arias\inst{1} \and%
  Rémy Haemmerlé\inst{1} \and \\ %
  Manuel V. Hermenegildo\inst{1,2} \and%
  José F. Morales\inst{2}}

\institute{%
  Universidad Politécnica de Madrid \and 
  IMDEA Software Institute 
}

\begin{document}
\maketitle

\begin{abstract}
  We present a new free library for Constraint Logic Programming over
  Finite Domains, included with the Ciao Prolog system.
  The library is entirely written in Prolog, leveraging on Ciao's
  module system and code transformation capabilities in order to
  achieve a highly modular design without compromising performance.
  We describe the interface, implementation, and design rationale of
  each modular component.
  The library meets several design goals: a high level of modularity,
  allowing the individual components to be replaced by different
  versions; high-efficiency, being competitive with other \fd
  implementations; a glass-box approach, so the user can specify new
  constraints at different levels; and a Prolog implementation, in order
  to ease the integration with Ciao's code analysis components.
  The core is built upon two small libraries which implement integer
  ranges and closures. On top of that, a \emph{finite domain variable}
  datatype is defined, taking care of constraint reexecution depending
  on range changes. These three libraries form what we call the
  \emph{\fd kernel} of the library.
  This \fd kernel is used in turn to implement several higher-level
  finite domain constraints, specified using indexicals. Together with
  a labeling module this layer forms what we name \emph{the \fd
    solver}.
  A final level integrates the \clpfd paradigm with
  our \fd solver. This is achieved using attributed variables and a
  compiler from the \clpfd language to the set of constraints
  provided by the solver.
  It should be noted that the user of the library is encouraged to
  work in any of those levels as seen convenient: from writing a
  new range module to enriching the set of \fd constraints by writing
  new indexicals.
\end{abstract}

\section{Introduction}
\label{sec:introduction}

Constraint Logic Programming (CLP)~\cite{survey94-short} is a natural
and well understood extension of Logic
Programming (LP) in which term unification is replaced by constraint
solving over a specific domain.  This brings a number of theoretical
and practical advantages which include increased expressive power and
declarativeness, as well as higher performance for certain application
domains.  The resulting CLP languages allow applying efficient,
incremental constraint solving techniques to a variety of problems in
a very natural way: constraint solving blends in elegantly with the
search facilities and the ability to represent partially determined
data that are inherent to logic programming.  As a result, many modern
Prolog systems offer different constraint solving capabilities.

One of the most successful instances of CLP is the class of constraint
logic languages using \emph{Finite Domains} (\fd).
Finite domains refer to those constraint systems in which constraint
variables can take values out of a finite set, typically of integers
(i.e., a {\em range}). 
They are very useful in a wide variety of problems, and thus many
Prolog systems offering constraint solving capabilities include a
finite domain solver.
In such systems, domain (range) definition
constraints as well as integer arithmetic and comparison constraints are
provided in order to specify problems.

Since the seminal paper of Van Hentenryck et al.~\cite{ccfd-jlp}, many
FD solvers adopt the so-called ``glass-box'' approach. Our FD Kernel also 
follows this approach, based on a unique primitive called an
\emph{indexical}. High-level constraints are then built/defined in terms of
primitive constraints.
An indexical has the form
\verb|X in r|, where $r$ is a range expression (defined in
\cref{fig:range:syntax}).  Intuitively, \verb|X in r| constrains the
\fd term (\fd variable or integer) \verb|X| to belong to the range
denoted by the term \verb|r|. In the definition of the range special
expressions are allowed. In particular, the expressions \verb|max(Y)|
and \verb|max(Y)| evaluate to the minimum and the maximum of the range
of the \fd variable \verb|Y|, and the expression \verb|dom(Y)|
evaluates to the current domain of \verb|Y|.
Constrains are solved partially in an incremental using consistency
techniques~\cite{Dib10arc_consistency_survey} which maintain the constraint network in
some coherent state (depending on the arc-consistency algorithm
used). This is done by monotone domain shrinking and propagation. When
all constraints are placed and all values have been propagated a call
is typically made to a \emph{labeling} predicate which performs an
enumeration-based search for sets of compatible instantiations for
each of the variables that remain not bound to a single value. We
refer to \cite{ccfd-jlp} for more details regarding indexicals and
finite domain constraint solving.

In this paper, we present a new free library for Constraint Logic
Programming over Finite Domains, included with the Ciao Prolog
system~\cite{hermenegildo11:ciao-design-tplp}.
  The library is entirely written in Prolog, leveraging on Ciao's
  module system and code transformation capabilities in order to
  achieve a highly modular design without compromising performance.
  We describe the interface, implementation, and design rationale of
  each modular component.
  The library meets several design goals: a high level of modularity,
  allowing the individual components to be replaced by different
  versions; high-efficiency, being competitive with other \fd
  implementations; a glass-box approach, so the user can specify new
  constraints at different levels; and a Prolog implementation, in order
  to ease the integration with Ciao's code analysis components.
  The core is built upon two small libraries which implement integer
  ranges and closures. On top of that, a \emph{finite domain variable}
  datatype is defined, taking care of constraint reexecution depending
  on range changes. These three libraries form what we call the
  \emph{\fd kernel} of the library.
  This \fd kernel is used in turn to implement several higher-level
  finite domain constraints, specified using indexicals. Together with
  a labeling module this layer forms what we name the \emph{\fd
    solver}.
  A final level integrates the \clpfd paradigm with
  our \fd solver. This is achieved using attributed variables and a
  compiler from the \clpfd language to the set of constraints
  provided by the solver.
  It should be noted that the user of the library is encouraged to
  work in any of those levels as seen convenient: from writing a
  new range module to enriching the set of \fd constraints by writing
  new indexicals.

\medskip

One of the first \clpfd implementations is the CHIP
system~\cite{chip}.  This commercial system follows a typical
black-box approach: it consists of a complete solver written in C and
interfaces in an opaque manner to a Prolog engine. This makes it
difficult for the programmer to understand what is happening in the
core of the system. Also, no facilities are provided for tweaking the
solver algorithms for a specific application.

More recent \clpfd systems such as those in
SICStus~\cite{Carlsson:1997:OFD:646452.692956}, GNU
Prolog~\cite{diaz11:gnu_prolog,clpfd}, and B-Prolog
\cite{Zhou:2006:PFC:1180162.1180163} are built instead following more
the glass-box approach. The basic constraints are decomposed into
smaller but highly optimized primitives (typically indexicals).
Consequently, the programmer has more latitude to extend the
constraints as needed. However, even if such systems can be easily
modified/extended at the interface level (e.g., both SICStus and
B-Prolog provide way to define new global constraints) they are much
harder to modify at the implementation level (e.g., it is not possible
to replace the implementation of range).

The Ciao \clpfd library that we present has more similarities with the
one recently developed for SWI
Prolog~\cite{springerlink:10.1007/978-3-642-29822-6_24}. Both are
fully written in Prolog and support unbound ranges. The SWI library is
clearly more complete than Ciao's (e.g., it provides some global
constraints and always terminating propagation), but it is designed in
a monolithic way: it is implemented in a single file, mixing different
language extensions (using classical Prolog \verb!term_expansion!
mechanisms) while the Ciao library is split in more around 20 modules
with a clear separation of the different language
extensions~\cite{composing-extensions-lopstr-11}.

Summarizing, our library differs in a number of ways from other
existing approaches:
\begin{itemize}
\item First, along with more recent libraries it differs from early
  systems in that it is written entirely in Prolog. This dispenses
  with the need for a foreign interface and opens up more
  opportunities for automatic program transformation and analysis.
  The use of the meta-predicates \verb!setarg/3! and \verb!call/1!
  means that the use of Prolog has a minimal impact on performance.
\item Second, the library is designed as a set of separate
  modules. This allows replacing a performance-critical part --- like the
  range code --- with a new implementation better suited for it.
\item Third, the library supports the ``glass-box'' approach fully,
  encouraging the user to access directly the low-level layers for
  performance-critical code without losing the convenience of the
  high-level CLP paradigm. Again, the fact that the implementation is
  fully in Prolog is the main enabler of this feature.
\item Lastly, we have prioritized extensibility, ease of modification,
  and flexibility, rather than micro-optimizations and pure raw
  speed. However, we argue that our design will accommodate several
  key optimizations like the ones of~\cite{wam-fd-iclp-diaz} without
  needing to extend the underlying WAM.
\end{itemize}

The rest of the paper proceeds as follows. In Sec.~\ref{sec:structure}
we present the architecture of the library and the interface of the
modules. In Sec.~\ref{sec:glass-box-effect} we discuss with an example
how to use the glass box approach at different levels for better
efficiency in a particular problem, with preliminary benchmarks
illustrating the gains. Finally, in Sec.~\ref{sec:conclusions} we
conclude and discuss related and future work.
\section{Architecture of the Ciao \clpfd Library}
\label{sec:structure}

The Ciao \clpfd library consists of seven modules grouped into three
logical layers plus two specialized Prolog to Prolog translators. In
the definition of these modules and interfaces we profit from Ciao's
module system~\cite{ciao-modules-cl2000} and Ciao's support for
assertions~\cite{ciaopp-sas03-journal-scp,hermenegildo11:ciao-design-tplp},
so that every predicate is correctly annotated with its types and
other relevant interface-related characteristics, as well as
documentation.  The translators are built using the Ciao
\emph{packages} mechanism~\cite{ciao-modules-cl2000}, which provides
integrated and modular support for syntax modification and code
transformations.
A description of the user interface for the library along with
up-to-date documentation may be found in the relevant part of the Ciao
manual.

\subsection{The Global Architecture}

The global architecture is illustrated in
Fig.~\ref{fig:global-architecture}. The kernel layer provides
facilities for range handling and propagation chains, which are used
for defining finite domain variables --- which, as mentioned before,
are different from the standard logical variables. The \fd layer
defines a finite set of constraints such as \verb!a+b=c/3!, using
indexicals. These constraints are translated form their indexical form
to a set of instructions of the kernel layer. Labeling and
branch-and-bound optimization search modules complete the finite
domain solver.

The \clpfd constraints are translated to \fd constraints by a \clpfd
compiler. We use attributed variables to attach a finite domain
variable to every logical variable involved in \clpfd
constraints. Thus, the \clpfd layer is thin and of very low overhead.

\begin{figure}[t]
  \centering
\begin{tikzpicture}[mbox/.style={draw, text centered, rounded
    corners=5pt, inner sep=6pt}, marrow/.style={->,>=latex}]
\node[mbox] (fd_var) at (0.5, 1) {\fd Term};
\node[mbox] (fd_propags) at (-1.5, -0.8) {Propagators};
\node[mbox] (fd_range) at (1.5, -0.8) {Range};
\draw[marrow] (fd_var) -- (fd_range);
\draw[marrow] (fd_var) -- (fd_propags);
\node[mbox, dotted, text width=5 cm, text height=3cm] (kernel) at (-0.25,0.3) {}; 
\node (N) at (-1.75,1.5) {\bf \fd Kernel};

\node[mbox] (fd_const) at (4, -1.25) [right] {\fd Constraints};
\node[mbox] (labeling) at (4,0.25) [right] {Labeling};
\node[mbox] (optim) at (4,1.75) [right] {B\&B Optimization};

\draw[marrow] (optim)  to [out=180,in=25] (kernel) ;
\draw[marrow] (optim) to [out=-45,in=45] (fd_const);
\draw[marrow] (labeling)  to [out=180,in=0] (kernel);
\draw[marrow] (labeling) to [out=-90,in=120]  (fd_const);
\draw[marrow] (fd_const)  to [out=180,in=-20] node[below=13,pos=0.6] {{\footnotesize Idx Compiler}} (kernel) ;

\node[mbox, dotted, text width=11 cm, text height=5cm] (solver) at (2.25,0) {}; 
\node (N) at (6.5,-2.25) {\bf \fd Solver};

\node[mbox] (clp_rt) at  (2.25,3.65) {\clpfd Run Time};
\draw[marrow] (clp_rt) -- (solver) node[midway,auto=left] {\clpfd Compiler};
\node[mbox, dotted, text width=12 cm, text height=7cm] (solver) at (2.25,0.5) {}; 
\node (N) at (-2.5,3.5) {\bf \clpfd};
\end{tikzpicture}
  \caption{The Ciao \clpfd Library Architecture.}
  \label{fig:global-architecture}
\end{figure}

\subsection{The Finite Domain Kernel}

The finite domain kernel is the most important part of the library.
Its implementation freely follows the design of the GNU~Prolog \fd
solver (\cite{clpfd} provides a general overview of this solver).
A finite domain variable is composed of a range and several
propagation chains. When the submission of a constraint modifies the
range of a finite domain variable, other finite domain variables
depending on that range are updated by firing up constraints stored in
propagation chains. The propagation events are executed in a
synchronous way, meaning that a range change will fail if any of its
dependent constraints cannot be satisfied.

The kernel implements arithmetic over ranges (pointwise operations,
union, intersection complementation, ...) and management of
propagation chains, amounting to the delay of Prolog goals on
arbitrary events. These two elements are used to implement the two
basic operations of a finite domain variable: \verb!tell! and
\verb!prune!. The first one attempts to constrain a variable into a
particular range, while the second one ({\tt prune/2}) removes a value
form the range of a variable.
The variable code inspects the new and old ranges and wakes up the
suspended goals on a given variable.

All the data structures are coded in an object-oriented style.
Efficient access and in-place update are implemented by using the
\verb!setarg/3! primitive. We took special care to use \verb!setarg/3!
in a \emph{safe} way to avoid undesired side effects, such as those
described by Tarau~\cite{binprolog2006}.

\subsubsection{Ranges.}
Range handling is one of the most important parts of the library,
given the high frequency of range operations. Indeed, the library
supports three implementations for ranges: the standard one using
lists of closed integer intervals; an implementation using lists of
open (i.e., unbounded) intervals; and a bit-based implementation which
despite allowing unbound ranges is more suitable for problems
dealing with small ranges.\footnote{The implementation of the
  bit-based range uses arbitrary precision integers plus three non-ISO
  predicates for computing the least and most significative bits, and
  the number of active bits in such
  integers.  
  We implemented these predicates in C.} %
Indeed, the user is encouraged to implement new range modules which
are better suited to some particular problems.

The interface that a range module must implement is split into two
parts. The first one, shown in Fig.~\ref{fig:range:syntax}, deals with
range creation and manipulation. Each of the operations defined in the
figure has a corresponding predicate. For instance, bounds addition
\verb!t+t! is implemented by the predicate \verb!bound_add/3!, and
similarly for the rest of the predicates. Note that it is a convention
of the interface that any operation that tries to create an empty
range will fail. This is better for efficiency and we found no
practical example yet where this would be inconvenient.

Fig.~\ref{fig:range:preds} lists the rest of the predicates that a
range implementation must provide. They are mainly used for obtaining
information about a range and are instrumental for the labeling
algorithms.
\begin{figure}[t]
  \centering
  \begin{tabular}{lllllll}
  \verb|r|  && $::=$ & $\quad$ & \verb|t .. t| & $\qquad$ & (interval) \\
&&& $\mid$ &  \verb|{t}| && (singleton) \\
&&& $\mid$ &  \verb|r \/ r| && (union) \\
&&& $\mid$ &  \verb|r /\ r| && (intersection) \\
&&& $\mid$ &  \verb|- r| && (complementation) \\
&&& $\mid$ &  \verb|r + n| && (pointwise addition) \\
&&& $\mid$ &  \verb|r - n| && (pointwise subtraction) \\
&&& $\mid$ &  \verb|r * n| && (pointwise multiplication)\\
\\
\verb|t|  && ::= && \verb|min(Y)|  && (minimum) \\
&&& $\mid$ &\verb|max(Y)| && (maximum) \\
&&& $\mid$ &\verb|dom(Y)|  && (domain) \\
&&& $\mid$ &\verb|val(Y)| && (value) \\
&&& $\mid$ &$\verb|t+t| \mid \verb|t-t| \mid \verb|t*t| \mid \dots $   && (arithmetic expression) \\
&&& $\mid$ &\verb|n|  && (bound) \\
\end{tabular}
  \caption{Range Interface, Part 1: Syntax.}
  \label{fig:range:syntax}
\end{figure}

\begin{figure}[b]
  \centering
  \begin{tabular}{l@{\hspace{0.5cm}}l}
    \verb!fd_range_bound_t/1! &	Type of a range bound.\\
    \verb!fd_range_t/1! & Type of a range object.\\
    \verb!is_singleton/1! & True if range is a singleton.\\
    \verb!singleton_to_bound/2! & Returns the value of a singleton range.\\
    \verb!size/2! & Number of elements in a range.\\
    \verb!get_domain/2! & List of elements in a range.\\
    \verb!enum/2! & Backtracks throughout all the elements in a range.\\
    \verb!bound_const/2! & Correspondence of indexical constants with bounds.\\
  \end{tabular}
  \caption{Range Interface, Part 2: Predicates.}
  \label{fig:range:preds}
\end{figure}

\subsubsection{Propagation Chains.}
Propagation chains are just lists of goals meant to be executed when a
change in the range of a \fd variable happens. The module defines a
\emph{propagation chain structure} that is simply a named set of
chains. We support in-place update for the structure, thus allowing
efficient update of the propagation chains used in the finite domain
variables. The interface of the propagation chain module is presented
in Fig.~\ref{fig:pchains}.
\begin{figure}[t]
  \centering
  \begin{tabular}{l@{\hspace{0.5cm}}l}
    \verb!fd_pchains_t/1! & Type of a chain structure.\\
    \verb!fd_pchain_type_t/1! & Name of a chain.\\
    \verb!empty/1! & Returns an empty chain structure.\\
    \verb!add/3! & Adds a goal to a given chain.\\
    \verb!execute/2! & Wakes up a particular chain.\\
  \end{tabular}
  \caption{Propagation Chain Interface.}
  \label{fig:pchains}
\end{figure}
We use internal facilities of the Ciao module system in order to
efficiently implement \verb!execute/2!.

\subsubsection{Finite Domain Variables.}

An \fd variable is a structure consisting of a range and a propagation
chain.

In the current implementation, integers are considered to be finite
domain variables too. However, we are in the process of phasing out
this optimization as we incorporate more information into finite
domain variables to aid optimizations.

\fd variables are never unified, i.e., they cannot be substituted by
others or by integer values as is typically done by the Prolog
unification mechanism.  A priori, such variables have no
correspondence to Prolog logical variables.

Apart from accessing its range and propagation chain, the most
important operations that a finite domain variable supports is the
tell operation, which tries to update the \fd variable to a new range:
\begin{lstlisting}
tell_range(FdVar, TellRange):-
	fd_var:get_range(FdVar, VarRange),
        fd_range:intersect(VarRange, TellRange, NewRange),
        set_range_and_propagate(FdVar, VarRange, NewRange)
\end{lstlisting}
The propagation predicate will set the new range for the variable and
compare the new range with the old one. The current definition ---
following \cite{wam-fd-iclp-diaz} ---
supports four propagation events, depending on the range change:\\

\begin{tabular}{lll}
\textbf{dom:} & & The range changed. \\
\textbf{max:} & & The maximum of the range has changed. \\
\textbf{min:} & & The minimum of the range has changed. \\
\textbf{val:} & & The new range is a singleton. \\
\end{tabular}

\subsection{The Finite Domain Solver}

Once the finite domain kernel is in place, the finite domain solver is
just the labeling algorithm and a set of constraints defined using the
kernel. As mentioned before, the constraints are defined using
indexicals, of the form \verb!X in Range!. Such indexicals are
compiled to programs of the \fd kernel in a transparent way for the
user. The compilation is carried out by Ciao's source-to-source
transformation capabilities, which means that an input Prolog file
using the \verb!indexicals! package is processed in such a way that
predicates containing indexical definitions are replaced by their
compiled form.

The indexical syntax is intended to be compatible with syntax used in
SICStus and GNU Prolog. However, the use of Ciao's package system
means that the user may freely mix indexicals with Prolog code (or
with many other syntax extensions, such as, e.g., functional notation)
without any ill effect, as seen in
Appendix~\ref{sec:queens-example-using}.

\subsubsection{The Constraints Library.}
A reasonable set of local constraints is provided, covering most
examples that we have tried to date. We use the convention of using
\verb!t! for ground terms, such that in the constraint
\verb!'a+b<>c'/3!, all three arguments are assumed to be \fd
variables, whereas in the constraint named \verb!a+t<>c/3!, the second
argument is assumed to be an immutable singleton, and thus no
propagation chains will be installed on it.

\subsubsection{Labeling and Optimization Searches,}

This layer includes also typical labeling algorithms and branch and
bound optimization searches. %
In fact, the current labeling engine is a slight adaptation of the one
in the SWI \clpfd library: we opted for replacing the preliminary
version of the engine with this one from SWI, because of its many
useful features and easy adaptability to our library.\footnote{Some
  features of this engine are currently disabled, but we are planning
  to activate all such features shortly. The labeling engine was in
  fact extracted from the \emph{tor} library~\cite{tor2012}, where it
  is isolated in a single file.}
The porting task was relatively easy because the labeling engine is a
quite peripheral part of the library (i.e., it has very few code
dependencies).  It also underlines the high modularity of our library,
since two versions of the labeling are in fact available~\footnote{The
  old labeling engine can be found in revisions older than 14721 of
  Ciao 1.15.}.  Finally we obtained for free a common user interface
with SWI (and Yap).

The optimization searches uses a branch-and-bound algorithm with
restart to find a value that minimizes (or maximizes) the \fd
variable according the execution of a Prolog goal.  It offers a 
user-interface similar to the one provided by GNU Prolog.

\subsection{\clpfd}
With the \fd solver in place, supporting the \clpfd paradigm is a
matter of performing two mappings: logical variables must be put in 
correspondence with \fd variables and \clpfd constraints must be
translated to \fd constraints.

\subsubsection{Variable Wrapping.}

For every logical variable to be involved in a \clpfd constraint we
will attach to it an attribute containing an \fd variable:
\begin{lstlisting}
wrapper(A, X):-	get_attr_local(A, X), !.
wrapper(A, X):- var(A), !, fd_term:new(X), put_attr_local(A, X).
wrapper(X, X):- integer(X), !.
\end{lstlisting}
Logical variables and finite domain variables may communicate in two
ways. In the first one, two logical variables may be unified, needing
to link their underlying finite domain variables. We implement this
communication using the \verb!unify_hook! attribute:
\begin{lstlisting}
attr_unify_hook(IdxVar, Other):-
        ( nonvar(Other) ->
            ( integer(Other) ->
                fd_constraints:'a=t'(IdxVar, Other)
            ;   clpfd_error(type_error(Other), '='/2)
            )
        ; get_attr_local(Other, IdxVar_) ->
            fd_constraints:'a=b'(IdxVar, IdxVar_)
        ; put_attr_local(Other, IdxVar)
        ).
\end{lstlisting}
We simply call the \fd constraints \verb!'a=b'/2! and \verb!'a=t'/2!.

The other form of communication is instantiation of the logical
variable when the corresponding finite domain one gets a singleton
range. We modify the \verb!wrapper! predicate to add an instantiation
goal to the val chain of freshly created \fd vars, i.e., we replace the
second clause within the definition of the wrapper by the following one:
\begin{lstlisting}
wrapper(A, X):- var(A), !, fd_term:new(X), put_attr_local(A, X),
        % Force instantiation of A when X represents an integer
	fd_term:add_propag(X, val, 'fd_term:integerize'(X, A)).
\end{lstlisting}
This small example points out the possibilities of our scheme beyond
the current use as a support for indexicals.

\subsubsection{Constraint Compilation.}

The \fd solver provides a finite set of \fd constraints, however, in
the \clpfd side we may encounter constraints such as:
\begin{lstlisting}
      A #= B + C + D + E
\end{lstlisting}
which should be linearized to
\begin{lstlisting}
      A1 #= D  + E,
      B1 #= B  + C,
      A  #= A1 + B1
\end{lstlisting}
and then wrapped to:\footnote{We profit here from Ciao's functional
  notation such that for {\tt p(X,Y)}, {\tt \textasciitilde p(X) } is
  handled syntactically like a function with return value {\tt Y}.}
\begin{lstlisting}
      'a=b+c'(~wrapper(A1), ~wrapper(D),  ~wrapper(E)),
      'a=b+c'(~wrapper(B1), ~wrapper(B),  ~wrapper(C)),
      'a=b+c'(~wrapper(A) , ~wrapper(A1), ~wrapper(B1))
\end{lstlisting}
We provide a small compiler which takes care of this process, along
with other features like compile-time integer detection.

\section{Glass-Box Programming}
\label{sec:glass-box-effect}
As previously stated, we encourage the use of a glass box approach
when programming with this library. We will use the classical
queens program in order to illustrate some of the possibilities that
the library offers:
\begin{itemize}
\item The use of different range implementations.
\item The direct use of the \fd constraints, skipping the \clpfd compiler.
\item The definition of new \fd constraints using indexicals.
\item The definition of new atomic constraints directly using the
  solver kernel, thus skipping the indexical compiler.
\end{itemize}

\paragraph{Benchmarking Conditions:} We provide for illustration
purposes some preliminary experimental results. However, it is
important to point out that the library is not yet in a state in which
relevant absolute performance numbers can be produced and its
performance potential fully assessed, since it is still missing
important optimizations. Also, only two benchmarks are used.

The benchmarks were run using Ciao 1.15 (revision 14744) on an
Intel(R) Core(TM)2 CPU T7200 @ 2.00GHz computer. For reference, we
include also the corresponding numbers for SWI Prolog (v. 5.10.4). The
purpose is not to make an extensive comparison\footnote{This is left
  as future work where, in addition to implementing the optimizations
  mentioned, we will include comparison with a number of other systems
  as well.} but rather to have a simple, well understood baseline with
which to compare. 
We should note that we did not explore SWI's support
for custom constraints.  At the same time, during these tests we have
determined that backtracking over changes made by \verb!setarg/3! is
currently significantly slower in Ciao than in SWI, which, given the
reliance of the implementation on \verb!setarg/3!  gives us a clear
avenue for performance improvement, independently of any changes to
the library itself.

The complete program used in the benchmark is shown in
Appendix~\ref{sec:queens-example-using}. Basically a benchmark has
three run time parameters, the number of queens (n=N), the labeling
strategy (either ``step'' or ``first fail''\footnote{Comparing the Ciao
  and SWI libraries using the heuristic labeling strategies as ``first
  fail'' is relevant since both use the same code for labeling.}), and
the constraints used, whose meaning will be explained later. For SWI,
only the first two parameters carry significance.

\subsection{Range Implementations}
As previously stated, the library provides three range
implementations, selectable at compile-time. The standard one is
called ``Closed,'' and represents ranges using a Prolog list of
intervals of integers. Thus, every \fd variable is always
bound. ``Open'' is a variation of this approach where the intervals
are enriched with constants \verb!sup! and \verb!inf!. This imposes a
penalty on bound arithmetic. Lastly, we compare both against a simple
bit-vector implementation, done mostly in Prolog with a small support
from C. The results can be seen in Fig.~\ref{fig:queens-range}.
\begin{figure}[t]
  \centering
  \begin{tabular}{|l@{\hspace{0.2cm}}|l@{\hspace{0.3cm}}|l@{\hspace{0.2cm}}|l@{\hspace{0.3cm}}|l@{\hspace{0.3cm}}|}
    \hline
    Queens Parameters & Bits & Closed      & Open  & SWI  \\ \hline
    n=16, step, clpfd   &    0.916 & 1.144 & 1.432 & 1.050 \\ \hline
    n=16, step, fd      &    0.572 & 0.848 & 1.104 & --   \\\hline
    n=16, step, idx     &    0.388 & 0.648 & 0.916 & --   \\\hline
    n=16, step, kernel  &    0.224 & 0.336 & 0.368 & --   \\\hline
    n=90, ff, clpfd    &    2.080 & 2.052  & 2.484 & 1.071 \\\hline
    n=90, ff, fd       &    1.112 & 1.272  & 1.592 & --   \\\hline
    n=90, ff, idx      &    0.752 & 1.124  & 1.588 & --   \\\hline
    n=90, ff, kernel   &    0.388 & 0.408 & 0.432 & --   \\\hline
  \end{tabular}
  \caption{Queens Benchmark.}
\label{fig:queens-range}
\end{figure}
The differences go from negligible to more than 50\%. In a different
benchmark (bridge), the closed interval version was 25\% faster than
the open one.

\subsection{Constraint Implementations}
We now focus on the different possibilities that the library allows
for \fd constraint programming.

In the queens program, the main constraint of the problem is expressed
by the \verb|diff/3| constraint:
\begin{lstlisting}
diff(X, Y, I) :-
    X   #\= Y,
    X   #\= Y+I,
    X+I #\= Y.
\end{lstlisting}
where \verb!I! will be always an integer.

However, the compiler cannot (yet) detect that \verb!I! is an integer,
and may perform some unnecessary linearization. We may skip the
compiler and define \verb!diff! using directly the \fd constraints:
\begin{lstlisting}
diff(X, Y, I):-
    fd_constraints:'a<>b'(~w(X),~w(Y)),
    fd_constraints:'a<>b+t'(X, Y, I),
    fd_constraints:'a<>b+t'(Y, X, I).
\end{lstlisting}
The speedup is considerable, getting close to 50\% speedup in some
cases. Indeed, the compiler should be improved to produce this kind of
code by default.

The user may notice that the above three constraints may be encoded by
using just two indexicals. For instance one can use the following
definition for \verb|diff/3|:
\begin{lstlisting}
diff(X,Y,I):-
    idx_diff(~w(X), ~w(Y), I).
idx_diff(X, Y, I) +:
    X in -{val(Y), val(Y)+c(I), val(Y)-c(I)},
    Y in -{val(X), val(X)+c(I), val(X)-c(I)}.
\end{lstlisting}
Again, the improvement is up to 40\% from the previous version.

However, the constraint \verb!diff! can be improved significantly by using
directly the kernel delay mechanism (val chain) and \fd variable
operations. In particular, we use the optimized kernel \verb|prune/2|
operation that removes a single element form the range of a
variable:
\begin{lstlisting}
diff(X, Y, I):-
	wrapper(X, X0), wrapper(Y, Y0)
  	fd_term:add_propag(Y, val, 'queens:cstr'(X0, Y0, I)),
  	fd_term:add_propag(X, val, 'queens:cstr'(Y0, X0, I)).

% Y is always a singleton.
cstr(X, Y, I):-
  	fd_term:integerize(Y, Y0),
  	fd_term:prune(X, Y0),
  	Y1 is Y0 + I,
  	fd_term:prune(X, Y1),
  	Y2 is Y0 - I,
  	fd_term:prune(X, Y2).
\end{lstlisting}
We reach around 80\% speedup from the first version, and this result
is optimal regarding what the user can do. Additional speedups can be
achieved, but not without going beyond our glass-box approach.
Indeed, our \clpfd compiler is simpler given that we are working on a
new translator that directly generates custom kernel constraints from
\clpfd constraints.

\section{Conclusions and Future Work}
\label{sec:conclusions}

The Ciao \clpfd library described is distributed with the latest Ciao
version, available at \url{http://ciaohome.org}. Although included in
the main distribution, it lives in the \verb!contrib! directory, as it
should be considered at a beta stage of development.

Even if we did not include yet important optimizations that should
improve significantly the performance of the library, the current
results are encouraging. The library has been used successfully 
internally within the Ciao development team in a number of projects.

The modular design and low coupling of components allow their easy
replacement and improvement. Indeed, every individual piece may be
used in a glass-box fashion. We expect that the use of Prolog will
allow the integration with Ciao's powerful static analyzers. At the
same time, the clear separation of run-time and compile-time phases
allows the modification and the improvement of the translation schemes
in an independent manner.  Indeed, the advantages of this design have
already been showcased in~\cite{clp-to-js-iclp12}, where a Prolog to
Javascript cross-compiler was used to provide a JS version of the
library and which only required replacing a few lines of code.  Using
this cross-compiler \clpfd programs can be run on the server side or
on the browser side unchanged.

Regarding future work, we distinguish two main lines: the kernel and
the \clpfd compiler.

For the kernel, the first priority is to finish settling down its
interface. While we consider it mature, some optimizations --- like
avoiding reexecution --- may require that we include more information
in our \fd variable structure, range modification times, etc.  Indeed,
we would like to support more strategies for propagators than the
current linear one. Support for some global constraints is on the
roadmap, and will likely mean the addition of more propagation chains.

The library features primitive but very useful statistics. However we
think it is not enough and we are working on an \fd instrumentation
package that will provide detailed statistics and profiling. This is
key in order to extract the maximum performance from the library.
Once we get detailed profiling information from a wide variety of
benchmarks, a better range implementation will be due.

Regarding the \clpfd compiler, the current version should be
considered a proof of concept. Indeed, we are studying alternative
strategies including the generation of custom kernels or specialized
\fd constraints for each particular program in contrast to the current
approach of mapping a \clpfd program to a fixed set of primitive
constraints. CiaoPP --- Ciao's powerful abstract interpretation engine
--- could be used in the translation, providing information about the
\clpfd program to the \clpfd compiler so it can generate an optimal
kernel of \fd code for that program.  In this sense, we think that we
will follow the CiaoPP approach of combining inference with
user-provided annotations in the new \clpfd compiler.

\subsection*{Acknowledgments}

The authors would like to thank the anonymous reviewers for their
insightful comments.

The research leading to these results has received funding from the
Madrid Regional Government under CM project P2009/TIC/1465
(PROMETIDOS), and the Spanish Ministry of Economy and Competitiveness
under project TIN-2008-05624 {\em DOVES}.  The research by R\'emy
Haemmerl\'e has also been supported by PICD, the Programme for
Attracting Talent / young PHDs of the Montegancedo Campus of
International Excellence.

\appendix

\section{Complete Code for the Queens Example}
\label{sec:queens-example-using}
\small
\begin{lstlisting}
queens(N, L, Lab, Const) :-
	length(L, N),
	domain(L, 1, N),
	safe(L, Const),
	labeling(Lab, L).

safe([], _Const).
safe([X|L], Const) :-
	noattack(L, X, 1, Const),
	safe(L, Const).

noattack([], _, _, _Const).
noattack([Y|L], X, I, Const) :-
	diff(Const, X, Y, I),
	I1 is I + 1,
	noattack(L, X, I1, Const).

diff(clpfd, X, Y, I) :-
 	X #\= Y, X #\= Y+I, X+I #\= Y.

diff(fd, X,Y,I):-
	fd_diff(~wrapper(X), ~wrapper(Y), I).

fd_diff(X, Y, I):-
	fd_constraints:'a<>b'(X,Y),
	fd_constraints:'a<>b+t'(X,Y,I),
	fd_constraints:'a<>b+t'(Y,X,I).

diff(idx, X,Y,I):-
	idx_diff(~wrapper(X), ~wrapper(Y), I).

idx_diff(X, Y, I) +:
 	X in -{val(Y), val(Y)+c(I), val(Y)-c(I)},
 	Y in -{val(X), val(X)-c(I), val(X)+c(I)}.

diff(kernel, X,Y,I):-
	kernel_diff(~wrapper(X), ~wrapper(Y), I).

kernel_diff(X, Y, I) :-
  	fd_term:add_propag(Y, val, 'queens:cstr'(X, Y, I)),
  	fd_term:add_propag(X, val, 'queens:cstr'(Y, X, I)).

cstr(X, Y, I):-
  	fd_term:integerize(Y, Y0),
  	fd_term:prune(X, Y0),
  	Y1 is Y0 + I, fd_term:prune(X, Y1),
  	Y2 is Y0 - I, fd_term:prune(X, Y2).
\end{lstlisting}

\end{document}